\documentclass[conference,comsoc]{IEEEtran}
\usepackage[T1]{fontenc}
\usepackage{amsmath}
\usepackage[cmintegrals]{newtxmath}
\usepackage{url}
\usepackage{soul,color,graphicx,float,epstopdf}
\usepackage{tablefootnote,textcomp,gensymb}
\usepackage{eurosym,booktabs,multirow}
\usepackage[caption=false]{subfig}
\usepackage{amsfonts} 
\usepackage{algorithmic}
\usepackage{array}
\usepackage{stfloats}
\usepackage{float}
\usepackage{mathtools}
\usepackage{graphicx}
\usepackage{pdfpages}
\usepackage{subfig} 
\usepackage{comment}
\usepackage{balance}
\usepackage{xcolor}
\usepackage{comment}
\usepackage{algorithm}
\usepackage[normalem]{ulem}
\usepackage[implicit=false]{hyperref}
\usepackage{enumitem}
\definecolor{Orange}{rgb}{1,0.5,0}
\newcommand{\todo}[1]{\textsf{\textbf{\textcolor{Orange}{[#1]}}}}

\begin{document}
\title{Building the Self-Improvement Loop: Error Detection and Correction in Goal-Oriented Semantic Communications}

\author{\IEEEauthorblockN{
Peizheng Li,
Xinyi Lin,
Adnan Aijaz
}\\ 
\vspace{-3.00mm}
\IEEEauthorblockA{
Bristol Research and Innovation Laboratory, Toshiba Europe Ltd., U.K.\\
Email: {\{Peizheng.Li, Xinyi.Lin, Adnan.Aijaz\}@toshiba-bril.com}}}

\maketitle

\begin{abstract}
Error detection and correction are essential for ensuring robust and reliable operation in modern communication systems, particularly in complex transmission environments. However, discussions on these topics have largely been overlooked in semantic communication (SemCom), which focuses on transmitting meaning rather than symbols, leading to significant improvements in communication efficiency. Despite these advantages, semantic errors---stemming from discrepancies between transmitted and received meanings---present a major challenge to system reliability. This paper addresses this gap by proposing a comprehensive framework for detecting and correcting semantic errors in SemCom systems. We formally define semantic error, detection, and correction mechanisms, and identify key sources of semantic errors. To address these challenges, we develop a Gaussian process (GP)-based method for latent space monitoring to detect errors, alongside a human-in-the-loop reinforcement learning (HITL-RL) approach to optimize semantic model configurations using user feedback. Experimental results validate the effectiveness of the proposed methods in mitigating semantic errors under various conditions, including adversarial attacks, input feature changes, physical channel variations, and user preference shifts. 
This work lays the foundation for more reliable and adaptive SemCom systems with robust semantic error management techniques.

\end{abstract}


\section{Introduction}
\label{sec:introduction}
To address the growing demand for intelligent and efficient communication in 6G networks, and to surpass the limits of traditional Shannon capacity, goal-oriented semantic communication (SemCom) has emerged as a promising paradigm~\cite{qin2021semantic}. Unlike conventional communication systems, which generally aim to maximize data throughput and minimize errors in bit-level transmission, SemCom prioritizes the transmission of relevant, task-specific knowledge, reducing redundant or irrelevant information~\cite{luo2022semantic}. This data-driven approach can significantly enhance communication efficiency, especially in scenarios where shared knowledge between transmitter and receiver allows for context-aware encoding and decoding. SemCom is viewed as a key enabler for 6G network~\cite{li2023open,strinati2024goal}.

The promise of SemCom has recently attracted research interests from various perspectives. One popular research avenue involves the investigation of semantic compression strategies, which focus on identifying and interpreting only the essential information to convey meaning. The semantic compression technologies are developed for specific tasks such as text~\cite{li2024crossword}, image~\cite{luo2018deepsic}, and video~\cite{tian2023non}, which are particularly useful in scenarios like human-robot interaction, autonomous driving, or Internet of Things (IoT), where reducing the data load while retaining critical meaning can lead to significant performance gains in terms of power consumption, bandwidth, and latency. Another trendy research area is semantic transmission, which leverages the fact that not all bits in a message are equally important for understanding the underlying meaning. Particularly, techniques such as joint source-channel coding (JSCC) are proposed to optimize the end-to-end communication process~\cite{park2024joint,tung2023deep,weng2021semantic}.

Despite extensive research focused on SemCom, most existing literature assumes that SemCom system is primarily unidirectional without either the error handling ability or a control loop with feedback. The semantic error, defined as the discrepancy between the transmitted and recovered information, has received less attention. Unlike physical and medium access control (MAC) layers, where techniques such as forward error correction (FEC) and automatic repeat request (ARQ) are proposed for reliable data transmission, the virtual semantic layer lacks a robust error management strategy. 
We believe detecting and correcting semantic errors are crucial in a SemCom system for several reasons: \emph{(1)} the detection and correction processes preserve the intended meaning by ensuring that the sender’s message is accurately conveyed, maintaining contextual accuracy to prevent misinterpretation of context-sensitive information; \emph{(2)} these mechanisms optimize resource utilization by focusing on transmitting meaning rather than ensuring perfect data accuracy; \emph{(3)} they enhance human-machine interaction by enabling systems to better understand user intent, even in the presence of input imperfections; and \emph{(4)} robust communication is ensured by providing resilience against noise and distortions that could compromise the conveyed meaning.
While previous studies~\cite{mridha2019semantic, naz2024optimizing, wu2024intelligent} have explored semantic error detection and correction, they primarily address mismatches at the terminal level without incorporating autonomous intelligence. A comprehensive framework for intelligent semantic error detection and correction applicable across the semantic channel remains absent.

Motivated by the gap, in this paper, we give a general definition of semantic error and develop a robust approach for identifying and addressing semantic errors. The main contributions of this paper are summarized as follows.
\begin{itemize}
    \item \textit{Semantic error definition and classification:} 
    To our knowledge, this is the first study to explore semantic errors over a virtual semantic channel with a focus on practical communication system integration. We define \textit{semantic error}, \textit{semantic error detection}, and \textit{correction}, and outline potential sources of these errors.
    
    \item \textit{Semantic error detection via latent space deviation:} We leverage the data distribution in the latent space to quantify the semantic error. A method for monitoring latent vectors using Gaussian Processes (GP) is proposed to detect errors in the SemCom framework. 
    \item \textit{HITL-RL for SemCom:} A human-in-the-loop reinforcement learning (HITL-RL) approach is introduced to further enhance the reliability of SemCom by leveraging human feedback. This feedback informs the training of the RL agent, optimizing hyperparameters for SemCom models and JSCC channels.
    
    \item \textit{Verification of GP for SemCom:} We validate the effectiveness of the proposed GP-based error detection in a variational autoencoder (VAE)-implemented SemCom system with JSCC. The method is tested against three factors: input feature changes, physical channel variations, and adversarial attacks targeting the semantic encoder.
    
    \item \textit{HITL-RL validation in multi-user scenarios:} The HITL-RL method is evaluated in a multi-user JSCC broadcasting scenario. After training, the RL agent effectively configures the SemCom model and knowledge base (KB) by utilizing user feedback.
\end{itemize}

The rest of this paper is organized as follows. Sec.~\ref{sec:defining semantic error} defines error detection and correction in SemCom and outlines the root causes of semantic error. Sec.~\ref{sec:methods of semantic error} introduces latent space distribution monitoring and user feedback collection methods for detecting and correcting semantic errors in theory.
Sec.~\ref{sec:gp} describes the proposed GP-based approach for semantic error detection through latent space monitoring. Sec.~\ref{sec:HITL} details the integration of SemCom with HITL-RL to address semantic errors. Sec.~\ref{sec:simulation and results} presents the experimental setting and corresponding results of the GP and HITL-RL used for SemCom. In-depth discussions about the presented framework are presented in Sec.~\ref{sec:discussion}, and Sec.~\ref{sec:conclusion} concludes the paper.

\section{Error detection and correction in SemCom}
\label{sec:defining semantic error}
The foundation of SemCom is the framework of generation and translation of the semantic representation, which, from the data perspective, refers to the data’s compression and recovery; while from the implementation perspective, refers to the semantic encoder and decoder. 
The transmission of semantic information relies on physical channels, which are typically designed for bit-level accuracy and are capable of handling noise, distortion, and interference. In contrast, the semantic channel is a virtual concept that abstracts the transmission of meaning in SemCom, prioritizing semantic integrity over raw data accuracy. Current research views the SemCom framework as unidirectional, where semantic information is extracted, transmitted, and then recovered sequentially.

\begin{figure}[t]
\centering
\includegraphics[scale=0.72]{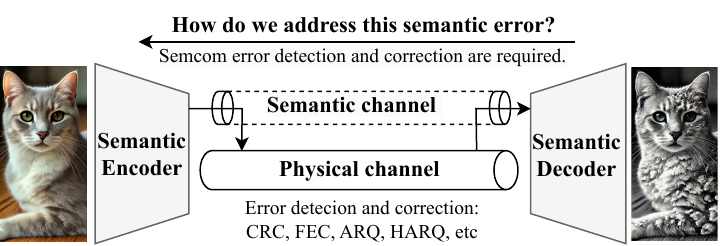}
\vspace{-5.45mm}
\caption{An illustration of the semantic error.}
\vspace{-1.45mm}
\label{fig:semcom error illustration}
\end{figure}

While the fidelity and accuracy of data recovery are key concerns in SemCom, most measures focus on the design of encoder-decoder models. From semantic channel perspective, however, there are no established mechanisms for semantic error detection and correction to improve the quality of the transmitted semantic information and, consequently, enhance the performance and reliability of SemCom. An example is shown in Fig.~\ref{fig:semcom error illustration}. Such considerations are crucial for the practical operation and integration of communication systems. 

Therefore, in this section, we begin by defining the concepts of semantic error, along with semantic error detection and correction. We then identify the potential root causes of semantic errors within the SemCom framework and outline corresponding solutions for addressing them.

\begin{itemize}[leftmargin=*]
    \item \textit{\textbf{Semantic Error:} a discrepancy between the transmitted raw information and the recovered information at the receiver in a goal-oriented semantic communication system.}

     \item \textit{\textbf{Semantic Error Detection:} the process of identifying semantic errors.}

    \item \textit{\textbf{Semantic Error Correction:} the process of correcting the transmitted semantic information according to the results of semantic error detection.}
\end{itemize}


Semantic errors primarily arise from a mismatch or misalignment between the prior domain knowledge embedded in the SemCom encoder/decoder and the new execution domain. The causes of this mismatch are varied, including:
\begin{enumerate}[leftmargin=*]
    \item \textbf{Input feature change:} This refers to the statistical properties of the input feature used to the ML-based semantic models changing over time. It is a significant challenge for SemCom models when the distribution of the feature in the inference stage is different from the prior knowledge learned from the training feature.
    
    \item \textbf{Channel noise and changes:} The SemCom model's performance deteriorates when the channel property changes significantly. That is because current SemCom models usually adopt the JSCC scheme to improve transmission efficiency. That is to say that the channel's prosperity was mesmerized by the SemCom encoder, so the obvious channel changes might lead to the misfunction of the encoder, and introduce the semantic error.
    
    \item \textbf{Adversarial attack targeting the NN-based encoder and decoder:} With adversarial attacks, carefully crafted imperceptible perturbations can be added to the input data of SemCom models. That can mislead the semantic neural network (NN) into making incorrect predictions, which in turn, generates the semantic error. 
    
    \item \textbf{User changes:} User preferences and tastes often change over time or differ across individuals. A static SemCom model cannot adapt to these variations, resulting in semantic errors from users' subjective perspectives.
\end{enumerate}

The first three factors are closely intertwined with the data and model, while the final factor is more subjective. Therefore, semantic error detection should employ distinct methods tailored to each specific factor.

\section{Semantic Error detection and correction methods}
\label{sec:methods of semantic error}
As SemCom sees broader use, semantic errors are likely to occur frequently, especially when encountering the factors mentioned above. Consequently, we believe that semantic error detection and correlation should serve as a core mechanism within the SemCom framework, enhancing its robustness in dynamic environments and with diverse users, ensuring correct and satisfactory operation.
Within current SemCom concepts, semantic information is considered a compact representation of raw data, with its precise measurements or meanings being agnostic. This presents significant challenges for detecting and correcting semantic errors, as these errors cannot be easily identified through quantitative metrics at the receiver.

Hence, in this paper, we propose two novel methods for semantic error detection, respectively, that is:
\begin{enumerate}[leftmargin=*]
    \item \textbf{Leveraging the data distribution on the latent space.} The latent space of the SemCom model represents the compressed input data distribution. Compared with the raw input data, the dimension of the latent space is usually small and compact. The input distribution changes can be directly reflected in the latent space. Thus, the monitoring of the latent space distribution is a valuable way to identify the semantic error. Especially considering the practice of a communication system, the monitoring of the encoded data either in the sender or receiver is more practical and communication effective.
    
    \item \textbf{Through the UE's collective feedback collection.} This is particularly raised for the semantic unsatisfactory caused by the UE's change. In this case, the performance of SemCom is determined by the UE's subjective feeling. Therefore, we propose to collect feedback from UEs to help the performance evaluation of SemCom models. For instance, semantic errors that trigger negative responses from users can be detected. Inspired by the concept of HITL-RL, we propose the HITL-RL SemCom method for semantic error detection correction. 
\end{enumerate}

The semantic error correction process will then utilize the results from semantic error detection to update both the KB and the semantic model accordingly.



\section{Distribution deviation identification with Gaussian processes}
\label{sec:gp}
\begin{figure}[t]
\centering
\includegraphics[scale=0.78]{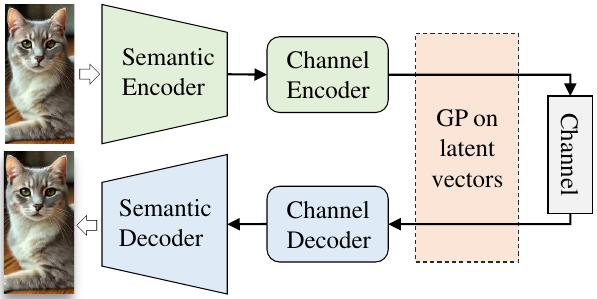}
\caption{Latent vector monitoring with GP for semantic error detection.}
\vspace{-1.45mm}
\label{fig:gp_semcom}
\end{figure}

As mentioned in the previous section, semantic errors can be identified by monitoring the latent space within the SemCom framework. The goal is to detect anomaly points in the latent space that indicate significant deviations from the learned data distribution. To effectively capture the complex, non-linear relationships in the latent space and account for uncertainty, we propose using GPs. GPs offer a probabilistic approach to modeling latent space distributions and quantifying uncertainty, making them useful for identifying outliers and anomalies. Fig.~\ref{fig:gp_semcom} illustrates the latent vector monitoring process using GP for the SemCom framework.

A GP defines a distribution over functions and is characterized by a mean function \( m(x) \) and a covariance function \( k(x, x') \). For a set of input latent variables \( Z = \{z_1, z_2, \dots, z_N\} \), the GP prior over the function values is defined as:
\begin{equation}
f(Z) \sim \mathcal{GP}(m(Z), k(Z, Z')),
\end{equation}
where \( k(Z, Z') \) is the covariance kernel that defines the relationship between different latent variables. The choice of kernel dictates the smoothness and general properties of the GP. A common choice for the covariance kernel is the radial basis function (RBF) kernel:

\begin{equation}
k(z, z') = \exp\left(-\frac{||z - z'||^2}{2\ell^2}\right),
\end{equation}
where $\ell$ is the length scale parameter.

The GP provides a non-parametric way to model the distribution of latent variables, capturing complex relationships in the data while offering uncertainty estimates. This is particularly useful for tasks like anomaly detection in the latent space, where the GP’s predictive uncertainty helps identify points that deviate from the expected behavior.

For a test point $z_*$, the GP provides a posterior predictive distribution:
\begin{equation}
p(f(z_*) | Z, f(Z)) = \mathcal{N}(\mu(z_*), \sigma^2(z_*)),
\end{equation}
where the mean $\mu(z_*)$ and variance $\sigma^2(z_*)$ are given by:
\begin{equation}
\begin{aligned}
\mu(z_*) &= k(z_*, Z)^T K(Z, Z)^{-1} f(Z), \\ 
\sigma^2(z_*) &= k(z_*, z_*) - k(z_*, Z)^T K(Z, Z)^{-1} k(z_*, Z).
\end{aligned}
\end{equation}
Here, $Z$ represents the matrix of training latent variables, $K(Z, Z)$ is the covariance matrix of the training points, and $k(z_*, Z)$ is the covariance between the test point and the training points.

The mean $\mu(z_*)$ represents the GP's prediction for the latent variable at $z_*$, while the variance $\sigma^2(z_*)$ quantifies the uncertainty of the prediction. Anomalies can be detected in two ways:
\begin{itemize}
    \item High uncertainty: If $\sigma^2(z_*)$ is large, it indicates that the test point lies far from the training data in the latent space, indicating an outlier or anomaly.
    \item Deviating mean: If the predicted latent variable $\mu(z_*)$ deviates significantly from the typical values observed during training, it could also signal an anomaly. This occurs when the test point is far from the expected behavior learned from the training data.
\end{itemize}

To quantify the likelihood of a point being an anomaly, an anomaly score can be defined as:
\begin{equation}
\text{Anomaly score}(z_*) = ||z_* - \mu(z_*)||^2+\sigma^2(z_*).
\label{eq:anomaly score}
\end{equation}
If this score exceeds a certain threshold, the point is classified as an anomaly.

\section{HITL-RL for Semantic error detection and correction}
\label{sec:HITL}
\begin{figure}[t]
\centering
\includegraphics[scale=0.78]{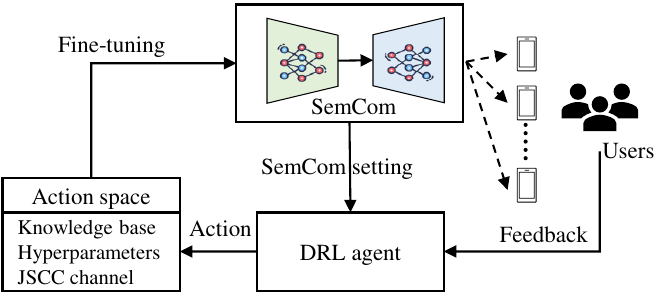}
\caption{Semcom model adjust via HITL-RL.}
\vspace{-1.45mm}
\label{fig:rlhf_semcom}
\end{figure}
Semantic errors arising from changes in human objective values cannot be addressed by the GP method, as the data distribution in the latent space remains unchanged. To manage this type of error, the SemCom framework must incorporate direct user feedback. By collecting feedback on user satisfaction with the current SemCom service, the system can proactively adjust and update its models. This includes managing and updating the KB, tuning NN model hyperparameters, and addressing channel noise during JSCC training.

Therefore, we propose to use HITL-RL in the SemCom to \emph{(1)} keep the users of SemCom service actively and continuously involved during the learning and updating process; \emph{(2)} enable the RL agent to make certain decisions of the SemCom model re-training/tuning with these humans value.

In HITL-RL, the human interactions are integrated into the standard RL process to guide the agent's exploration and policy updates. 
In a typical RL setting, an agent interacts with an environment, modeled as a Markov Decision Process (MDP) defined by the tuple \( \langle S, A, P, R, \gamma \rangle \), where \( S \) is the state space, \( A \) is the action space, \( P(s'|s,a) \) is the state transition probability, \( R(s,a) \) is the reward function and \( \gamma \in [0,1] \) is the discount factor. The goal of the agent is to learn a policy \( \pi(a|s) \), which maximizes the expected cumulative reward~\cite{li2022rlops}:
\begin{equation}    
J(\pi) = \mathbb{E}_{\pi} \left[ \sum_{t=0}^{\infty} \gamma^t R(s_t, a_t) \right]
\end{equation}

In HITL-RL, human feedback modifies the reward signal or policy to improve learning. Let \( f_h(s_t, a_t) \) represent the human feedback function. The modified reward function \( \tilde{R}(s_t, a_t) \) becomes a combination of environment rewards and human feedback:
\begin{equation}    
\tilde{R}(s_t, a_t) = R(s_t, a_t) + \alpha f_h(s_t, a_t)
\label{eq:HITL-RL}
\end{equation}
where \( \alpha \) is a weighting factor determining the influence of human feedback.

The overall workflow of HITL-RL for SemCom can be described in the following steps:
\begin{enumerate}
    \item The agent observes the state \( s_t \) and selects an action \( a_t^{\text{agent}} \) according to its policy \( \pi_\theta(a|s) \) to change the SemCom models training hyperparameters and re-train/fine-tune the SemCom model.
    \item The agent updates its action to \( \tilde{a_t} \) based on human intervention (if applicable).
    \item The reward \( \tilde{R}(s_t, a_t) \) is computed using both environmental (SemCom setting) rewards and human feedback.
    \item The policy \( \pi_\theta \) is updated using the modified reward and the standard policy gradient method.
\end{enumerate}

\section{Simulation and results}
\label{sec:simulation and results}
\begin{figure*}[t]
\centering
\includegraphics[scale=0.75]{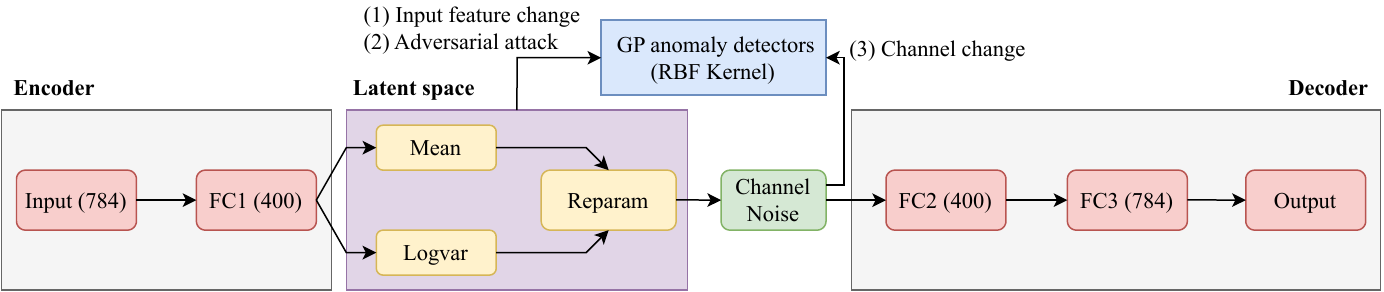}
\caption{Latent vector monitoring with GP for three types of error sources.}
\vspace{-1.45mm}
\label{fig:vae_gp_semcom_JSCC}
\end{figure*}

\subsection{Experimental setup}
\subsubsection{VAE for SemCom}
We consider the VAE architecture as the backbone of SemCom implementation. In VAE, the observed data $x \in \mathbb{R}^d$ is generated from a lower-dimensional latent variable $z \in \mathbb{R}^k$, where $k \ll d$. The relationship between the data and the latent variables is modeled by a generative process. The objective of VAE training is to learn an approximate posterior distribution over latent variables, denoted as \( q(z|x) \), which encodes the input data into a latent space. Due to its efficiency in capturing latent representations, the VAE is widely adopted in SemCom architectures. The VAE loss function is composed of two key terms:
\begin{equation}
\mathcal{L}(x) = \mathbb{E}_{q(z|x)}[\log p(x|z)] - D_{\text{KL}}\bigl(q(z|x) || p(z)\bigr),    
\label{eq:vae loss}
\end{equation}
where the first term represents the reconstruction loss, and the second term is the Kullback-Leibler (KL) divergence, which regularizes the approximate posterior \( q(z|x) \) to remain close to the prior distribution \( p(z) \). A comprehensive explanation of the VAE generative process can be found in~\cite{kingma2019introduction}.

\subsubsection{JSCC}
In default, we adopt the JSCC in the simulations of this paper. JSCC has been proposed in~\cite{bourtsoulatze2019deep}, which aims to eliminate the traditional separation of source and channel coding, while optimizing them simultaneously in SemCom. Let \( X \in \mathcal{X} \) represent the source information, and \( Y \in \mathcal{Y} \) represent the received signal at the destination. In JSCC, the traditional Shannon source and channel separation are integrated by designing a mapping \( f_{\theta}: \mathcal{X} \to \mathcal{S} \) that directly maps the source \( X \) to a signal \( S \), and a decoder \( g_{\phi}: \mathcal{S} \to \mathcal{Y} \), where:
\begin{equation}
S = f_{\theta}(X), \quad Y = g_{\phi}(S + N),
\end{equation}
where \( N \sim \mathcal{N}(0, \sigma^2) \) represents the additive white Gaussian noise (AWGN) in the channel.

\subsubsection{Adversarial attack}
To implement the adversarial attack on the VAE-based SemCom model, we use the fast gradient sign method (FGSM) algorithm. This method performs a single step of perturbation by taking a step of fixed length \( \epsilon \) in the direction of the gradient of the VAE loss function (Eq.~\ref{eq:vae loss}) with respect to the input data. The perturbation can be calculated as:

\begin{equation}
\eta = \epsilon \cdot \text{sign}(\nabla_x L(\theta, x, y)),
\label{eq:fgsm}
\end{equation}
where \( \epsilon \) represents the magnitude of the perturbation, controlling the strength of the adversarial attack, \( \nabla_x \) denotes the gradient of the loss \( L(\theta, x, y) \) with respect to the input \( x \), and \( y \) is the original label. The perturbed input \( x' \) is then given by $x' = x + \eta$.
FGSM creates adversarial examples by slightly altering the input in a way that maximizes the loss.



\subsubsection{Dataset}
We conducted experiments using the MNIST dataset.
The VAE encodes images into a 20-dimensional latent space with a hidden layer of 400 neurons. JSCC is applied by adding AWGN noise to the latent variables during transmission, with noise levels \(\sigma = 0.1\) to \(0.5\). The VAE was trained using the Adam optimizer with a learning rate of \(10^{-3}\) for 50 epochs. The model was evaluated on its reconstruction accuracy under different scenarios.

\subsection{GP for semantic error detection and correction}
As discussed in Sec.~\ref{sec:gp}, GP can be employed for latent space monitoring to detect semantic errors caused by input feature changes, channel noise and variations, as well as adversarial attacks targeting the semantic encoder/decoder. To validate the effectiveness of GP, we conducted three separate experiments.
Each GP was trained on the latent space using an RBF kernel. For each test sample's latent representation, the GP computes the predictive mean and variance, from which anomaly scores are derived.

\begin{figure*}[t]   
    \subfloat[\label{fig:roc}]{
      \begin{minipage}[t]{0.25\linewidth}
        \centering 
        \includegraphics[width=1.7in]{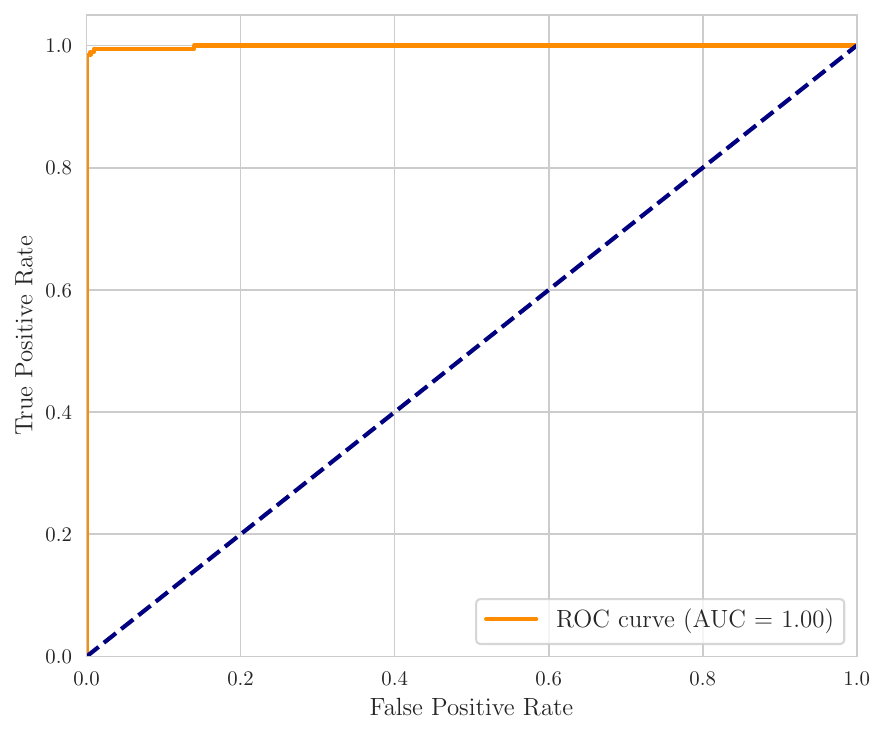}   
      \end{minipage}%
      }
        \subfloat[\label{fig:tSNE}]{
      \begin{minipage}[t]{0.25\linewidth}   
        \centering   
        \includegraphics[width=1.7in]{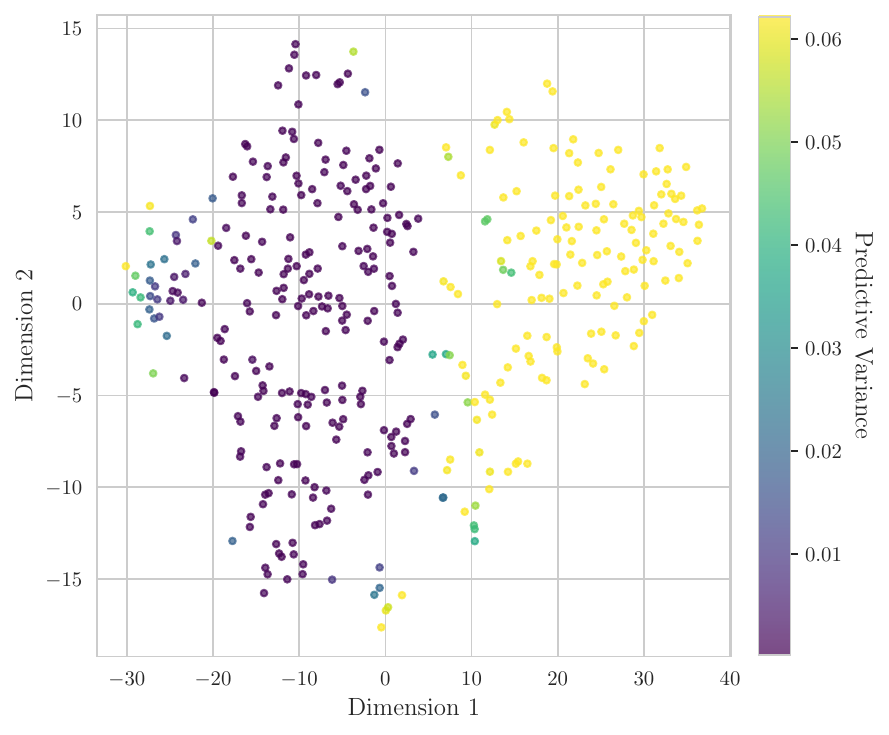}
      \end{minipage} 
      }
        \subfloat[\label{fig:confusion_matrix}]{
      \begin{minipage}[t]{0.25\linewidth}   
        \centering   
        \includegraphics[width=1.7in]{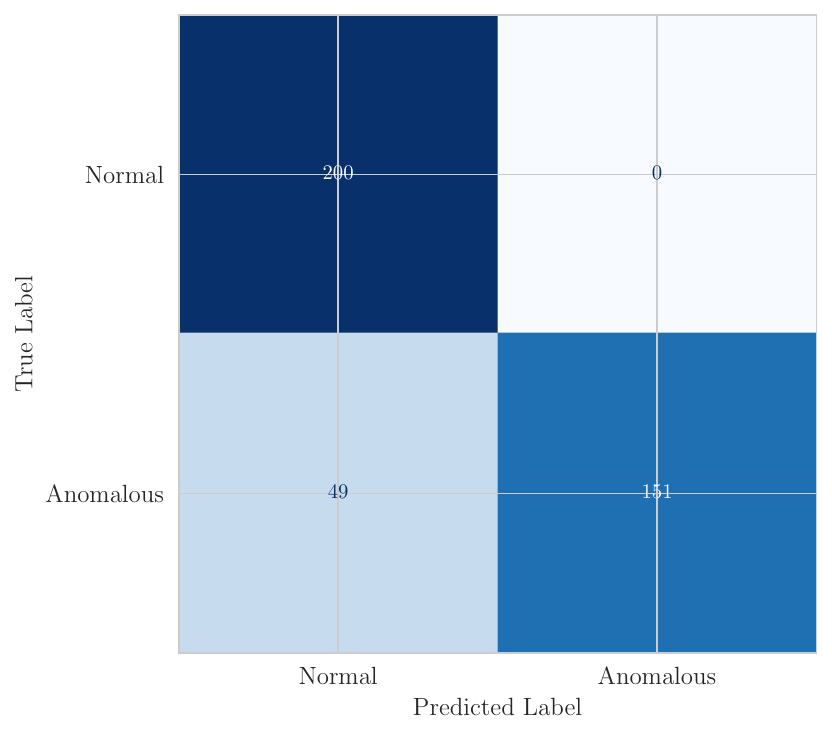}
      \end{minipage} 
      }
    \subfloat[\label{fig:reconstruct}]{
      \begin{minipage}[t]{0.25\linewidth}   
        \centering   
        \includegraphics[width=1.5in]{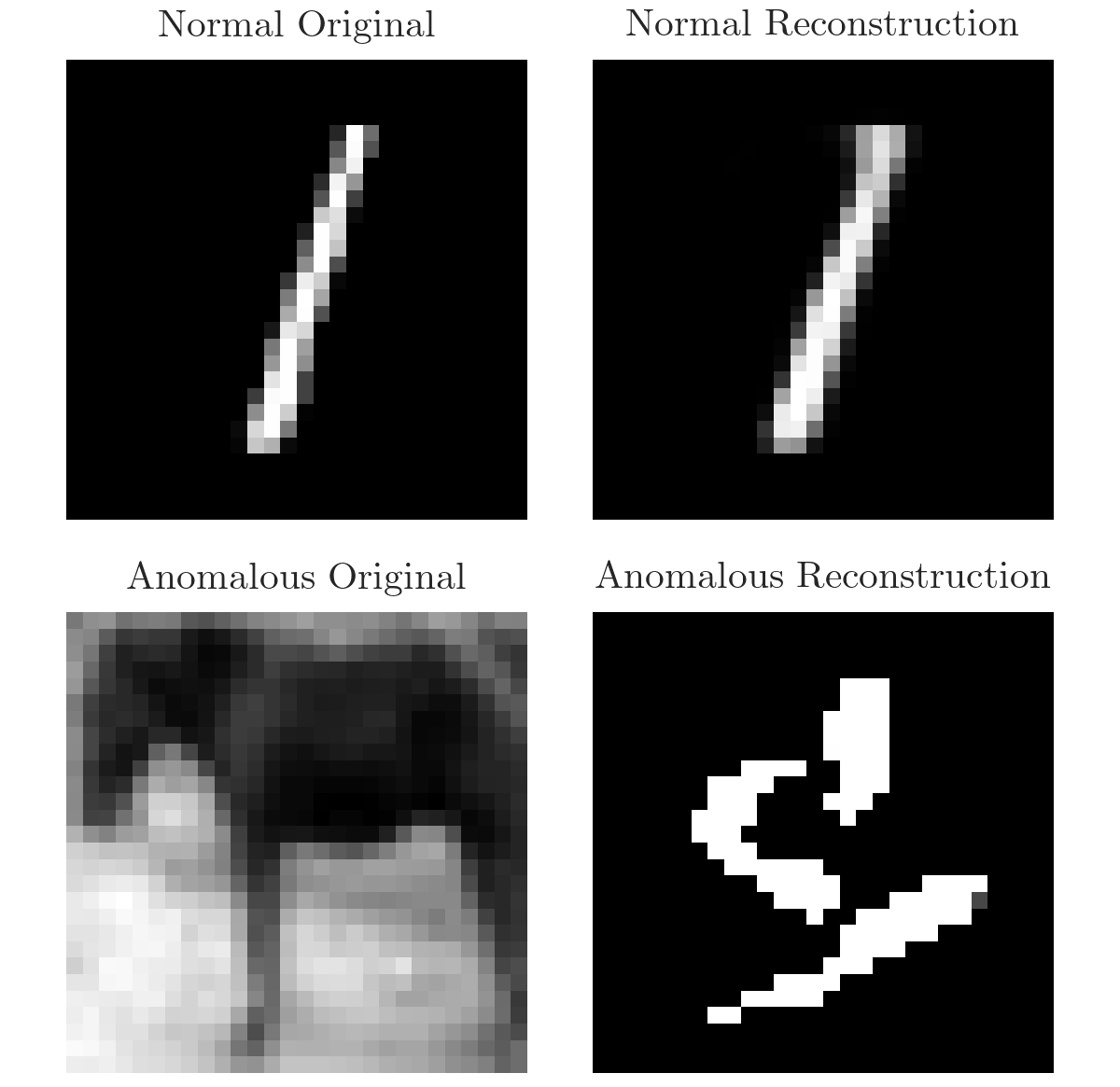}
      \end{minipage} 
      }
      \vspace{-1mm}
      \caption{
      GP-based anomaly detection for \textbf{input feature change}: (a) ROC curve; (b) t-SNE of latent space; (c) confusion matrix; (d) VAE reconstruction comparison (correct/incorrect input).
      } \label{fig:no_mnist_results}
\vspace{-1.45mm}
\end{figure*} 
\subsubsection{GP for input feature change detection}
We use the MNIST dataset for the VAE-based SemCom model training and GP training (on the latent variables $z$). However, we feed the CIFAR-10 dataset jointly with MNIST data into the trained model to mimic the change of input features during the deployment stage of the VAE model. In this process, we use the GP model to identify the abnormal distribution caused by the CIFAR-10 dataset in the latent space. In this process, the threshold of anomaly scores (Eq.~\ref{eq:anomaly score}) is set to 0.02. The anomaly detection results are shown in Fig.~\ref{fig:no_mnist_results}. 
It can be seen from the confusion matrix (Fig.~\ref{fig:confusion_matrix}) that the GP is able to identify most of the anomalous inputs from the normal inputs.

\subsubsection{GP for channel change detection}
\begin{figure*}[t]   
    \subfloat[\label{fig:ch_roc}]{
      \begin{minipage}[t]{0.25\linewidth}
        \centering 
        \includegraphics[width=1.75in]{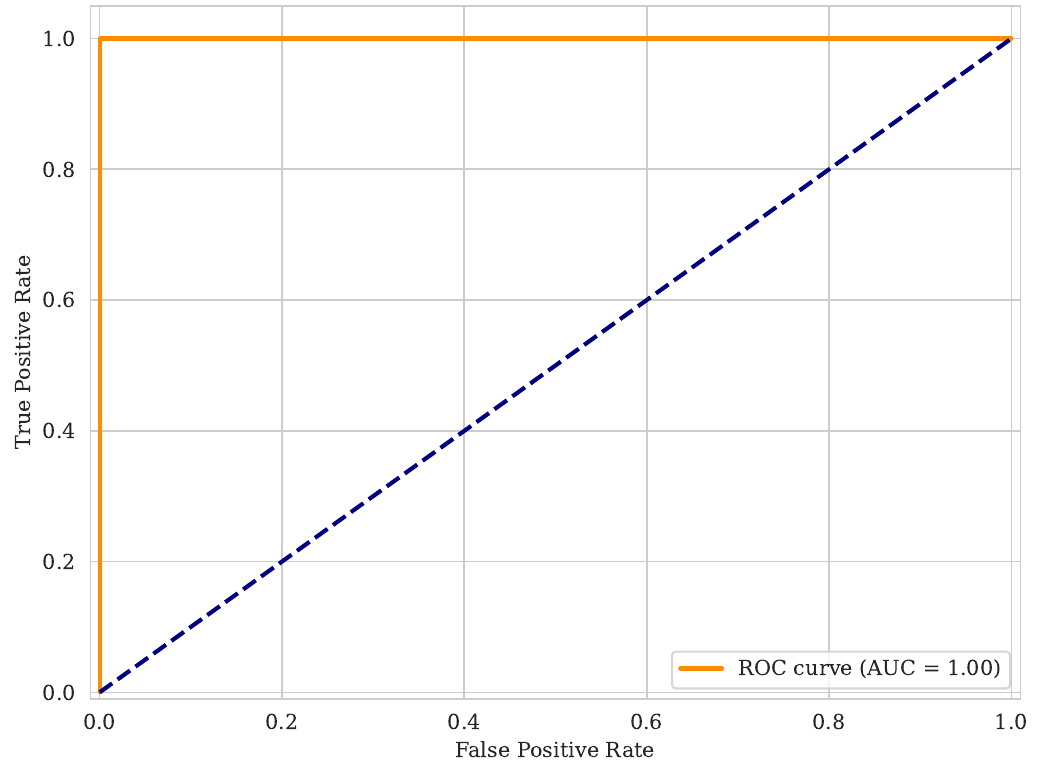}   
      \end{minipage}%
      }
        \subfloat[\label{fig:ch_tSNE}]{
      \begin{minipage}[t]{0.25\linewidth}   
        \centering   
        \includegraphics[width=1.7in]{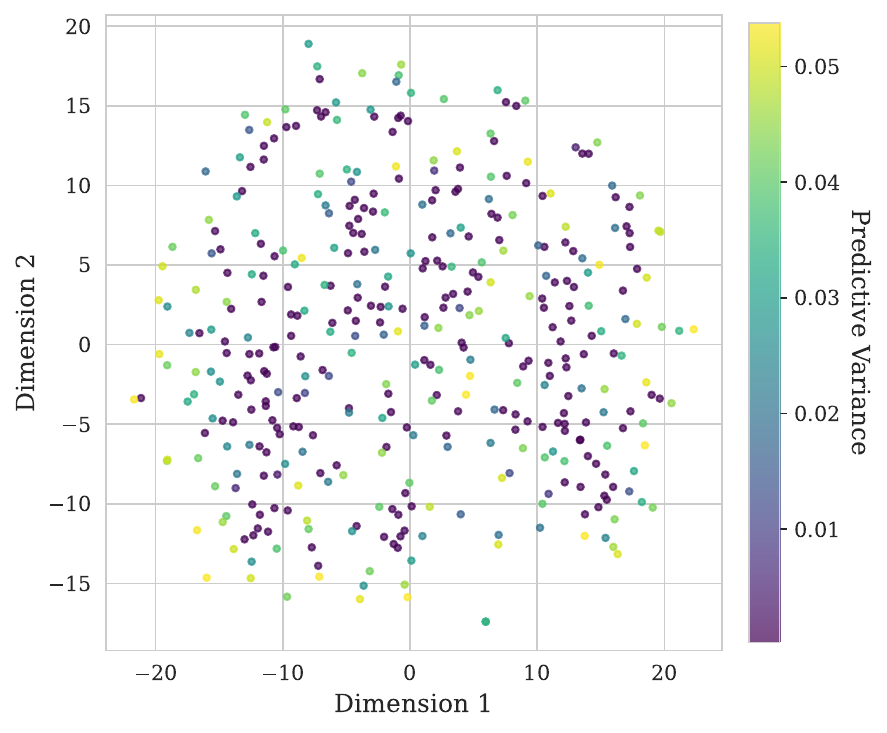}
      \end{minipage} 
      }
        \subfloat[\label{fig:ch_confusion_matrix}]{
      \begin{minipage}[t]{0.25\linewidth}   
        \centering   
        \includegraphics[width=1.75in]{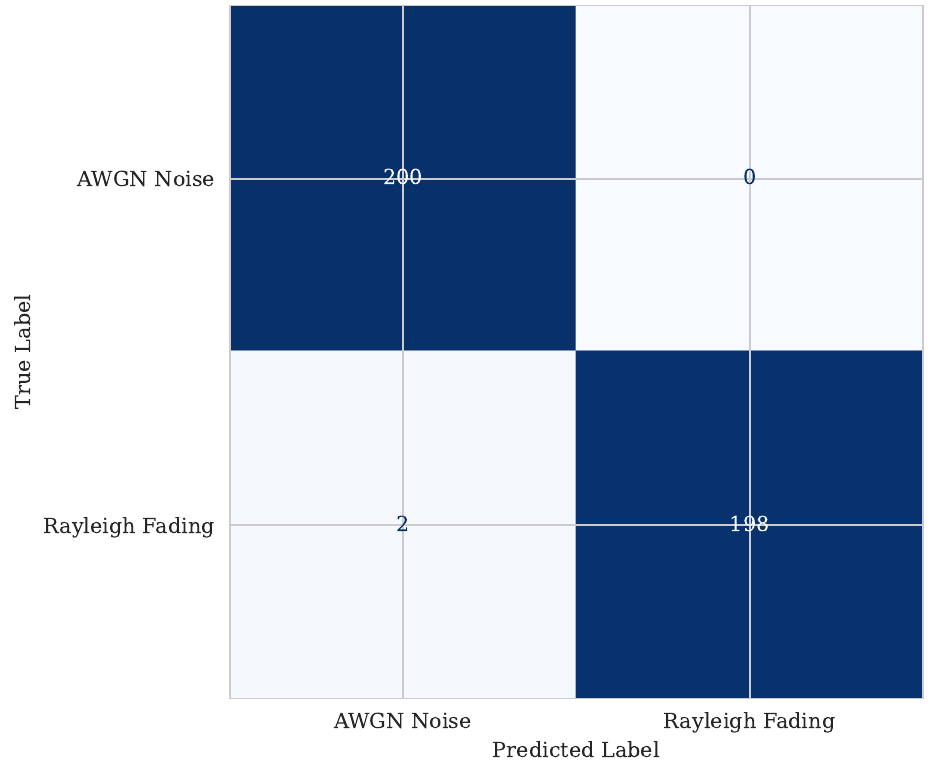}
      \end{minipage} 
      }
    \subfloat[\label{fig:channel_reconstruct}]{
      \begin{minipage}[t]{0.25\linewidth}   
        \centering   
        \includegraphics[width=1.5in]{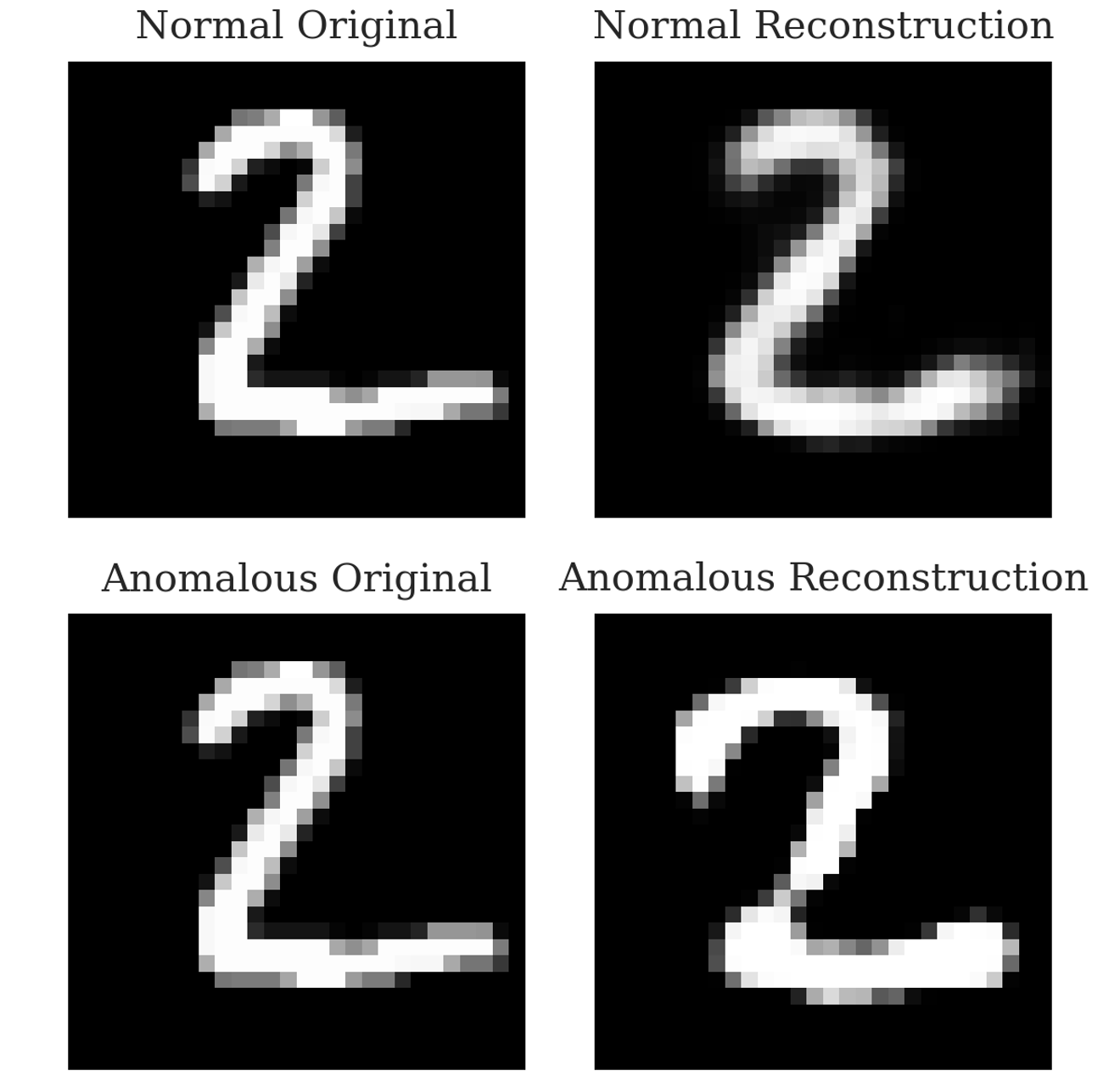}
      \end{minipage} 
      }
      \vspace{-1mm}
      \caption{GP anomaly detection in \textbf{JSCC with channel change}: (a) ROC curve; (b) t-SNE of latent space; (c) confusion matrix; (d) VAE reconstruction comparison (AWGN vs. Rayleigh fading).
      } \label{fig:ch_results}
\vspace{-1.45mm}
\end{figure*} 
We use the AWGN in JSCC to train the VAE model initially. 
However, we assume in its deployment stage, the channel is changed from the AWGN to a Rayleigh fading channel. Then the GP is used to detect the abnormal points in the latent space caused by the channel change. The detection results are illustrated in Fig.~\ref{fig:ch_results}. It can be seen that the change of channel features do have an impact on the reconstruction results of images (Fig.~\ref{fig:channel_reconstruct}), even if the distribution of normal/abnormal points has no significant deviation in the latent space, while the GP is able to detect the anomalous in the latent space.

\subsubsection{GP for adversarial attack detection}
\begin{figure*}[t]   
    \subfloat[\label{fig:aa_roc}]{
      \begin{minipage}[t]{0.25\linewidth}
        \centering 
        \includegraphics[width=1.7in]{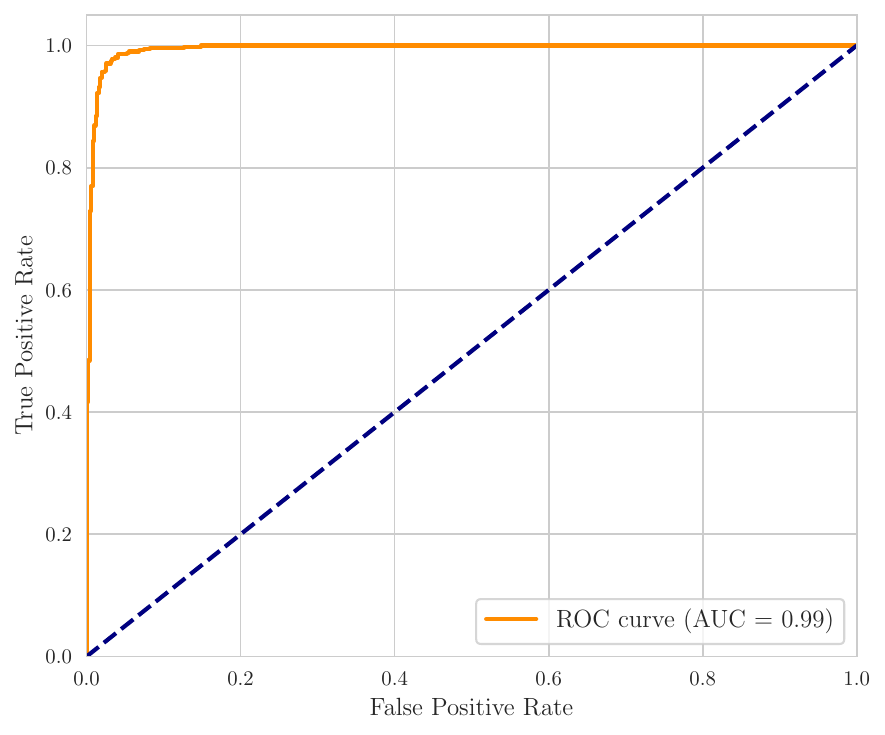}   
      \end{minipage}%
      }
        \subfloat[\label{fig:aa_tSNE}]{
      \begin{minipage}[t]{0.25\linewidth}   
        \centering   
        \includegraphics[width=1.7in]{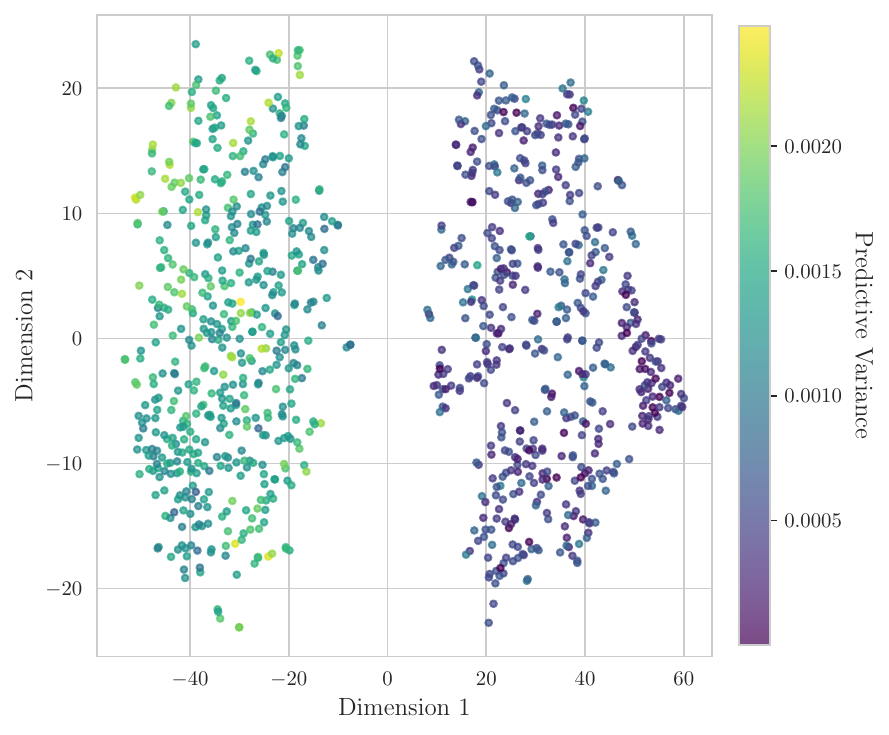}
      \end{minipage} 
      }
        \subfloat[\label{fig:aa_confusion_matrix}]{
      \begin{minipage}[t]{0.25\linewidth}   
        \centering   
        \includegraphics[width=1.7in]{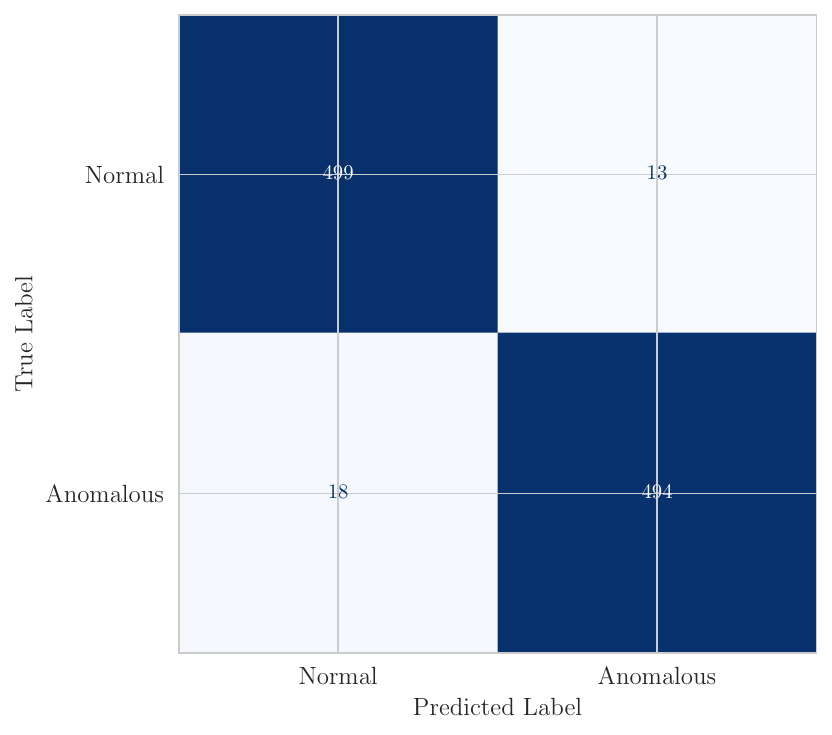}
      \end{minipage} 
      }
    \subfloat[\label{fig:aa_channel_reconstruct}]{
      \begin{minipage}[t]{0.25\linewidth}   
        \centering   
        \includegraphics[width=1.5in]{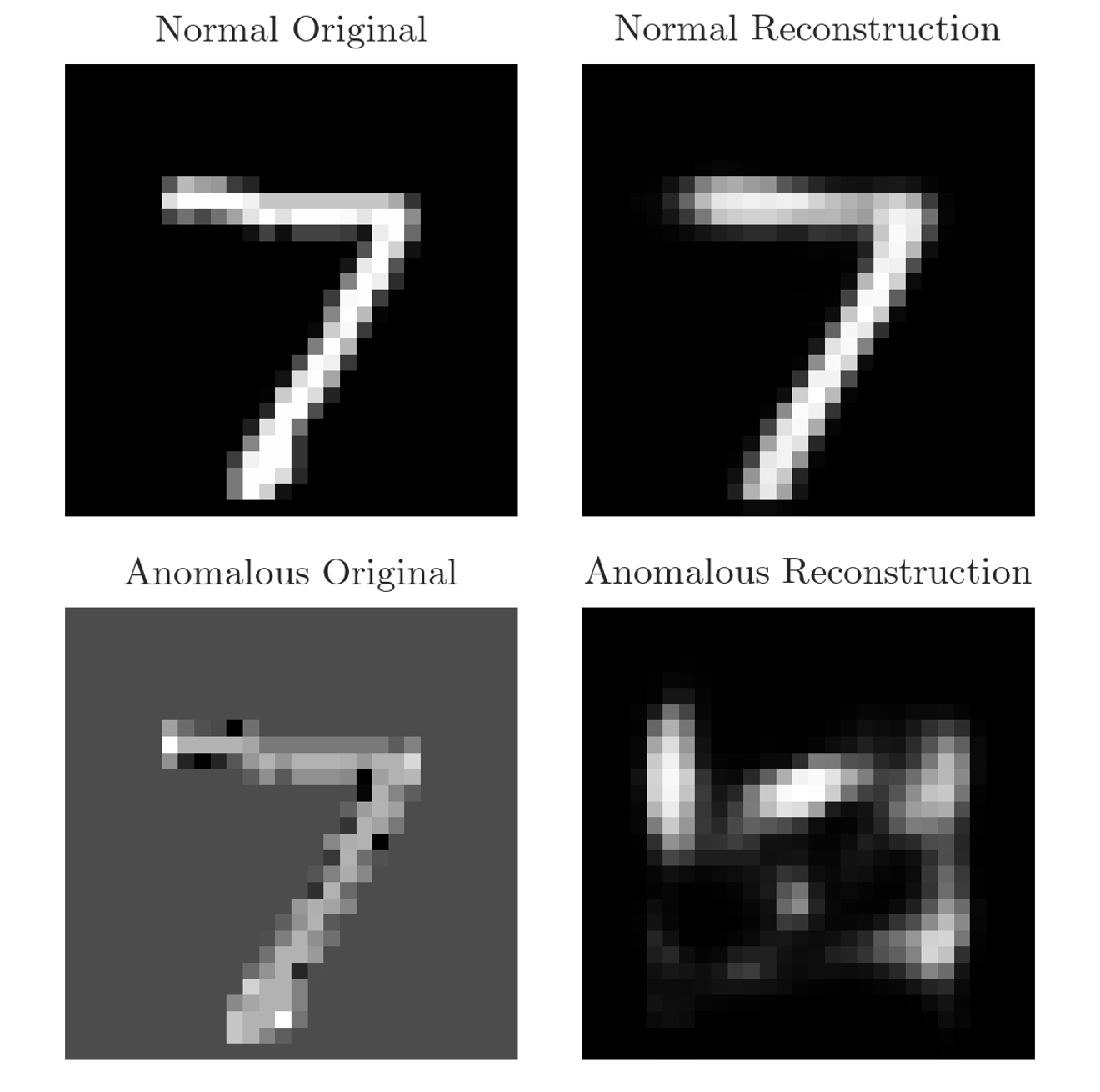}
      \end{minipage} 
      }
      \vspace{-1mm}
      \caption{GP anomaly detection for the \textbf{semantic encoder adversarial attacks}: (a) ROC curve; (b) t-SNE of latent space; (c) confusion matrix; (d) VAE reconstruction comparison (normal vs. adversarial noise perturbed input.
      } \label{fig:aa_results}
\vspace{-1.45mm}
\end{figure*} 
In this case, we still use the MNIST dataset to train the VAE model and GP is trained using the latent space distribution of VAE. However, we assume the adversarial attack is applied to the VAE (the semantic encoder more specifically) in the deployment stage using the FGSM algorithm. The trained GP model is leveraged to identify the adversarial anomalous. In this experiment, the $\epsilon$ in Eq.~\ref{eq:fgsm} is set to 0.3 for FGSM implementation. From the results shown in Fig.~\ref{fig:aa_results}, it can be seen that delicate designed adversarial noise has a huge impact on the VAE's image reconstruction, and a noise perturbed number ``7'' almost cannot be reconstructed by the VAE (Fig.~\ref{fig:aa_channel_reconstruct}). However, GP has a good performance in detecting the latent variables affected by the adversarial noise (Fig.~\ref{fig:aa_confusion_matrix}). It should be noted that in this case, the threshold of anomaly scores (Eq.~\ref{eq:anomaly score}) should be carefully selected, which is been set to 0.001 in this experiment.

\subsection{HITL-RL for semantic error detection and correction}
To verify the effectiveness of HITL-RL for semantic error detection and correction. We design a SemCom system that incorporates HITL-RL to optimize a VAE model in a multi-user JSCC broadcasting scenario. 

Specifically, we aim to improve the VAE's ability to reconstruct images under varying channel conditions and input characteristics by leveraging user feedback to dynamically adjust the training process.

In our experimental design, we consider a SemCom system where a transmitter communicates with five UE devices over Gaussian noise channels with different noise levels. We utilize the MNIST dataset for image data. During the initial training phase, the VAE model is trained exclusively on images of odd digits (labels 1, 3, 5, 7, 9) from the MNIST dataset, and we simulate an ideal channel with a noise standard deviation of $\sigma = 0$. This setup represents a scenario where the model has not been exposed to certain input features (even digits) or channel imperfections.

In the deployment phase, both odd and even digits are transmitted. Each UE experiences a different channel noise level, with standard deviations $\sigma_i \in {0.1, 0.2, 0.3, 0.4, 0.5}$ for $i = 1, 2, \dots, 5$, respectively. Upon receiving the transmitted data, each UE reconstructs the image using the VAE decoder and provides feedback in the form of the mean squared error (MSE) between the original and reconstructed images.

The experimental procedure involves a deep Q network (DQN) agent that interacts with the environment to minimize the overall reconstruction error. The \emph{state} at each time step $t$ is defined as $s_t = [\text{MSE}_1, \text{MSE}_2, \dots, \text{MSE}_5, \sigma_{\text{train}}, I_{\text{even}}]$, where $\text{MSE}_i$ is the average MSE for UE $i$, $\sigma_{\text{train}}$ is the training noise standard deviation, and $I_{\text{even}}$ is an indicator variable denoting whether even digits are included in the training data. The \emph{action} space consists of adjustments to the training process, including whether to include even digits, the choice of $\sigma_{\text{train}}$, the learning rate $\alpha$, and the number of training epochs $E$.

\begin{figure}[t]
\centering
\includegraphics[scale=0.5]{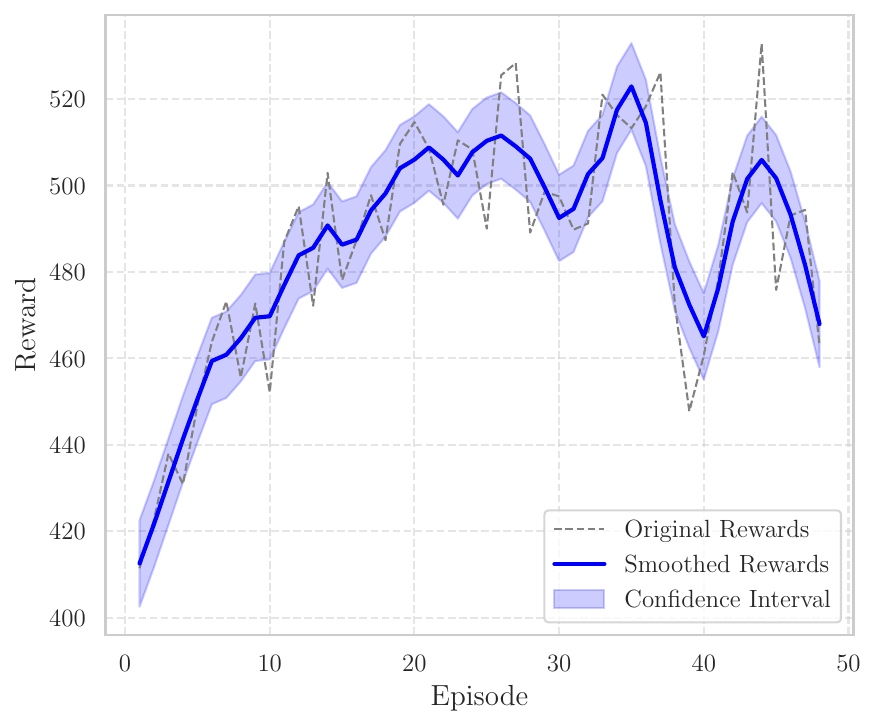}
\caption{HITL-RL training reward over episodes for the multi-user JSCC broadening SemCom system.}
\vspace{-1.45mm}
\label{fig:HITL_RL training}
\end{figure}

At each time step, the agent selects an action $a_t$ based on the current state $s_t$ using an $\epsilon$-greedy policy derived from the DQN. The VAE model is then fine-tuned according to the selected action parameters. The users' feedback is collected, and the next state $s_{t+1}$ is observed. According to Eq.~\ref{eq:HITL-RL}, the reward is computed as 
\begin{equation}
    r_t = \frac{1}{\overline{\text{MSE}} + \epsilon} - \alpha L_{\text{train}},
\end{equation}

where $\overline{\text{MSE}}$ represents the average MSE across all SemCom service users, $L_{\text{train}}$ denotes the training loss during fine-tuning, $\epsilon$ is a small constant to avoid division by zero, and $\alpha$ is a weighting factor to balance reconstruction accuracy and training efficiency.
The HITL-RL training curve is illustrated in Fig.~\ref{fig:HITL_RL training}. It can be observed that by iteratively updating the VAE model and the DQN agent through episodes consisting of time steps, the SemCom system learns to adapt to varying channel conditions and input characteristics.

\section{Discussions}
\label{sec:discussion}
In this work, GP is utilized to detect anomalies within the SemCom latent space, without considering sequential data over time. For future work, GP can be extended to model and predict changes in the latent space over time, providing a more dynamic approach to anomaly detection.

We use HIRL-RL to improve the reliability of the SemCom framework in the presence of semantic imperfections and errors. Primary results demonstrate the effectiveness of this approach; however, further qualitative research on this method is warranted. Additionally, the action space of HIRL-RL could be expanded to include direct operations on the KB and layered configurations of SemCom models.

Additionally, RL from human feedback (RLHF) presents a promising avenue for exploration, particularly when integrating the SemCom framework with large generative AI models. By first training a reward function based on human values---derived from the preferences of SemCom service UEs---this approach could guide the fine-tuning and updating of SemCom generative AI models.

\section{Conclusions}
\label{sec:conclusion}
This paper addresses the critical issue of semantic errors in goal-oriented SemCom systems by developing a comprehensive error detection and correction framework. We defined semantic errors, detection, and correction, and identified the root causes of these errors within practical communication scenarios. To effectively address semantic errors, we proposed a GP-based method for monitoring latent space deviations, as well as a HITL-RL approach to optimize semantic model parameters based on user feedback. Our experimental results demonstrated the robustness and adaptability of the proposed methods in handling semantic errors under various challenging conditions, including adversarial attacks, input feature changes, physical channel variations, and evolving user preferences. Future work may explore further optimizations and extensions of the framework to enhance its generalization and application in more diverse communication environments.

\section*{Acknowledgment}
This work was supported by the 6G-GOALS project under the 6G SNS-JU Horizon program, n.101139232.

\bibliographystyle{IEEEtran} %
\bibliography{IEEEabrv,references} 

\end{document}